# Performance Comparison of 112 Gb/s DMT, Nyquist PAM4 and Partial-Response PAM4 for Future 5G Ethernet-based Fronthaul Architecture

Nicklas Eiselt, *Student Member, IEEE,* Daniel Muench, *Member, IEEE,* Annika Dochhan, *Member, IEEE,* Helmut Griesser, *Member, IEEE,* Michael Eiselt, *Fellow, OSA, Senior Member, IEEE,* Juan José Vegas Olmos, *Senior Member, IEEE,* Idelfonso Tafur Monroy, *Senior Member, IEEE,* and Joerg-Peter Elbers, *Member, IEEE*

*Abstract*—For a future 5G Ethernet-based fronthaul architecture, 100G trunk lines of a transmission distance up to 10 km standard single mode fiber (SSMF) in combination with cheap grey optics to daisy chain cell site network interfaces are a promising cost- and power-efficient solution. For such a scenario, different intensity modulation and direct detect (IMDD) formats at a data rate of 112 Gb/s, namely Nyquist four-level pulse amplitude modulation (PAM4), discrete multi-tone transmission (DMT) and partial-response (PR) PAM4 are experimentally investigated, using a low-cost electro-absorption modulated laser (EML), a 25G driver and current state-of-the-art high speed 84 GS/s CMOS digital-to-analog converter (DAC) and analog-to-digital converter (ADC) test chips. Each modulation format is optimized independently for the desired scenario and their digital signal processing (DSP) requirements are investigated. The performance of Nyquist PAM4 and PR PAM4 depend very much on the efficiency of pre- and post-equalization. We show the necessity for at least 11 FFE-taps for pre-emphasis and up to 41 FFE coefficients at the receiver side. In addition, PR PAM4 requires an MLSE with four states to decode the signal back to a PAM4 signal. On the contrary, bit- and power-loading (BL, PL) is crucial for DMT and an FFT length of at least 512 is necessary. With optimized parameters, all modulation formats result in a very similar performances, demonstrating a transmission distance of up to 10 km over SSMF with bit error rates (BERs) below a FEC threshold of 4.4E-3, allowing error free transmission.

## I. INTRODUCTION

SEVERAL architectures for future 5G fronthaul networks are currently discussed. One potential approach is based on a wavelength division multiplexing (WDM) passive optical network (PON) [1], where multiple 10G feeds from large cell sites are transported to the central office without any statistical multiplexing. This is in particular interesting in environments

The results were obtained in the framework of the SENDATE Secure-DCI, ICirrus and SpeeD projects, partly funded by the German ministry of education and research (BMBF) under contracts 16KIS0477K and 13N1374 and by the European Commission under grant agreement No. 644526 and in the Marie Curie project ABACUS.

N. Eiselt is with the Department of Photonics Engineering, Technical University of Denmark and with ADVA Optical Networking SE in Meiningen, Germany (neiselt@advaoptical.com).
D. Muench, A. Dochhan, H. Griesser, M. H. Eiselt and J.-P. Elbers are with ADVA Optical Networking SE in Meiningen and Martinsried, Germany.
J. J. Vegas Olmos and I. Tafur Monroy are with the Department of Photonics Engineering, Technical University of Denmark.

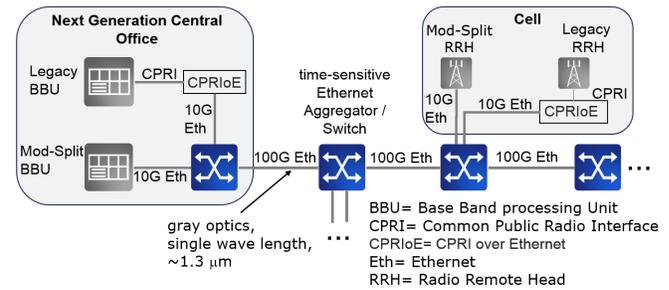

Fig. 1. Ethernet-based fronthaul architecture for 5G.

where fibers are a scarce resource or active equipment has to be avoided at intermediate sites. Another potential approach is based on Ethernet, which daisy chains cell site network interface devices with one or few 100G trunk lines and uses statistical multiplexing. Although this approach can be combined also with WDM, it is most beneficial in terms of power consumption, footprint and cost to transmit just a single wavelength in combination with cheap grey optics. As Ethernet data are transmitted, the required data rate is 103.125 Gb/s plus optional overhead. Fig. 1 shows the proposed Ethernet-based fronthaul architecture.

Particular benefits from using Ethernet are its widespreadness and ubiquity making Ethernet a cost-effective technology. Furthermore, it allows in principle all necessary topologies, like star-topologies for aggregating, tree topologies for adding and dropping of cells, and ring topologies for providing higher availability. In addition, it enables the reuse and adaptation of already available means for operation, administration and maintenance as well as for service level agreements or for self-optimizing networks. The aggregation and treatment of multiple traffic classes are also allowed. And finally, as a transparent transport layer it is agnostic to any new upcoming functional split between radio remote head (RRH) and baseband processing unit (BBU) [2]. Using CPRI (Common Public Radio Interface) over Ethernet (CPRIoE) mapper enables coexistence with legacy CPRI-based equipment [3].

All these benefits come at the price of stringent requirements for latency, latency variation and time synchronization. One approach to address the time sensitiveness is IEEE802.1CM, which discusses the application of existing means out of the toolbox for time-sensitive networking (IEEE802.1 TSN). Further constraints and assumptions for fronthaul networks are a topology with at least 6 hops or switches and at least 20 nodes [4]. These stringent requirements limit the maximum achievable distance of each 100G line. The allowed one-way latency is 75 $\mu s$ and bit error rates (BERs) of 1E-12 are required [5], [6].

Derived from the low-cost requirements are the considerations of using intensity modulation and direct detection (IMDD) together with grey optics for each 100G trunk line. To make the system as simple as possible, optical amplification and dispersion compensation are not an option. Consequently, the 1300 nm transmission window is required to prevent severe limitations from chromatic dispersion. For the modulation format, non-return to zero (NRZ) would be the simplest solution, requiring however expensive high-bandwidth optics and electronics. Hence, advanced modulation formats combined with DSP and FEC-encoding are considered, with potential candidates being PAM4 [7], [8], [9], duobinary/partial response PAM4 (PR PAM4) [10], [11], [12], DMT [13], [14] and carrier-less amplitude and phase modulation (CAP) [15]. Out of these, PAM4 and DMT have also been heavily discussed during the standardization activities of the IEEE802.3bs 400 GbE Task Force for next generation of intra-data center interconnects [16], resulting already in demonstrations of 100G real-time PHYs and offer good opportunities for reuse [17], [18].

A thorough comparison between the different modulation formats and their required DSP is essential in order to find the optimum solution for the scenario of interest. In this paper, we experimentally evaluate and compare Nyquist PAM4, PR PAM4 and DMT at a data rate of 112 Gb/s as potential modulation formats for the discussed transmission. To show its commercial feasibility, current state-of-the-art high speed CMOS DAC and ADC test chips with 84 GS/s are used together with a low-cost EML and a 25G driver. We demonstrate, that each format requires careful optimization, which at the end results in a very similar performance. A transmission distance of up to 10 km is successfully bridged.

The paper is structured as follows. Section II presents the experimental setup and the used components. The applied modulation formats and their required DSP are introduced in section III and IV. In section V the experimental results are demonstrated, discussed and compared. A conclusion is drawn in section VI.

## II. EXPERIMENTAL SETUP

Fig. 2 shows the employed experimental setup. Offline DSP is applied at the transmitter as well as at the receiver side and requires the use of a high-resolution DAC and ADC. Both operate at a sampling speed of 84 GS/s, have a nominal bit resolution of 8 bit and show a 3-dB bandwidth of around 15 GHz and 18 GHz, respectively. At the transmitter, the differential outputs of the DAC are amplified by a linear, differential input and single-ended output modulator driver amplifier (DA, MAOM-003115) driving directly the succeeding electro-absorption modulated laser (EML) [19]. This DA has integrated high frequency coils for biasing the EML, it provides an adjustable gain of up to 9 dB with a maximum output swing of 2 V and a 3-dB bandwidth of around 25 GHz. The EML operates at a fixed temperature of 45°C and the current of the distributed feedback laser (DFB) section is set to a maximum of 100 mA, which gives the highest linear range and the highest optical output power. At these operating conditions, the transmission wavelength of the EML is around 1308 nm and the 3-dB bandwidth is measured to be around 27 GHz but with a smooth roll-off. Fig. 3a) illustrates the unmodulated output spectrum of the EML, while Fig. 3b) shows the measured optical power vs. EML bias voltage. For transmission, a bias of -1.25 V was found to be optimum for all modulation formats and is therefore used throughout this paper.

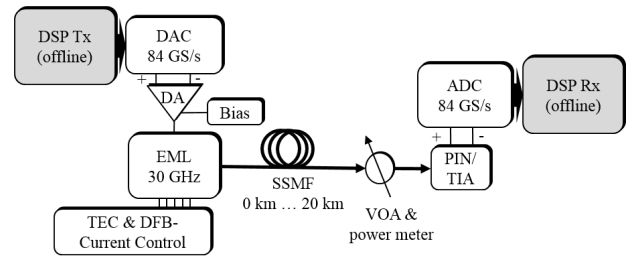

Fig. 2. Experimental transmission setup.

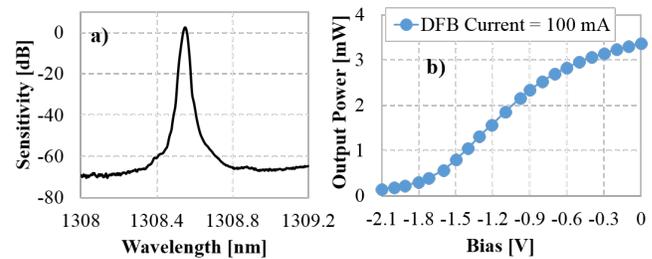

Fig. 3. Characteristics of the EML: a) measured output spectrum (unmodulated) and b) measured transfer function at a DFB-current of 100 mA and a temperature of 45°C.

The optical link setup consists of a conventional SSMF with an attenuation of around 0.32 dB/km at 1300 nm, a variable optical attenuator (VOA) with an integrated power monitor and a PIN-photodetector (Picometrix PT-40E) integrated with a linear trans-impedance amplifier (PIN/TIA) with a combined bandwidth of 35 GHz. Finally, the signal is sent back to the ADC and stored for offline processing. As the memory of this ADC is limited to 500000 samples, several blocks of data are used to make a valid statement of the bit error rate (BER) performance.

## III. MODULATION SCHEMES AND THEIR DSP

### A. Discrete Multi-tone Transmission

DMT as a special variant of orthogonal frequency division multiplexing (OFDM) employs the properties of Hermitian symmetry and the IFFT to create a real-valued signal with the frequency spectrum divided into orthogonal subcarriers. Each subcarrier can be modulated and the power of each subcarrier can be allocated based on the water filling method. This process is known as bit and power loading (BL, PL) and enables the effective compensation of channel impairments and component bandwidth limitations without applying complex signal processing, e.g. a simple 1-tap equalizer at the receiver side is efficient. To apply BL and PL, the transfer function of the transmission system is first estimated in terms of the signal-to-noise ratio (SNR) at the receiver with 16-QAM constellations with equal power on each subcarrier. Afterwards, Chow's margin-adaptive bit loading algorithm and Cioffi's power loading are applied to efficiently distribute the bits and allocate the power [20]. Fig. 5 shows the estimated SNR of the transmission setup for the optical back-to-back case, together with the corresponding bit and power allocation for a 112 Gb/s DMT signal. From the estimated SNR, we can also estimate the available bandwidth of the transmission system: an SNR of 15 dB or more is available up to 25 GHz, while it drops below 0 dBm for frequencies above 30 GHz. What is more, the EML exhibits a frequency drop at around 7 GHz in its transfer function, degrading the transmission performance. The clock-line of the DAC and ADC explains the frequency null at 21 GHz. Fig. 4 illustrates the DSP blocks of the analyzed DMT system and Table I summarizes the most important DMT system parameters. To meet the mentioned memory requirements of the DAC and the ADC, a DMT frame consists of 124 data symbols and four training symbols (in total 128 DMT symbols), which are used for channel estimation and synchronization.

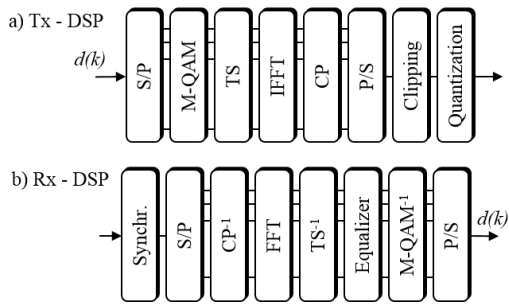

Fig. 4. DSP blocks of the implemented DMT system.

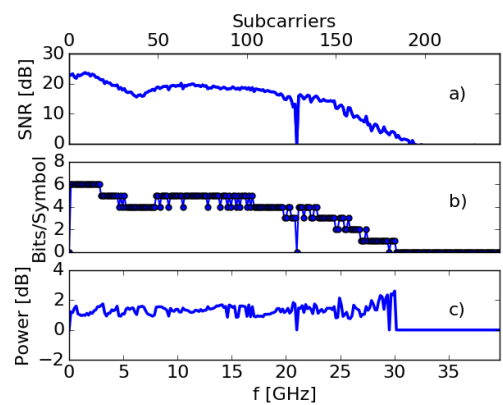

Fig. 5. a) Estimated SNR per subcarrier at the receiver for optical back-to-back and the corresponding b) bit and c) power loading.

### B. Nyquist PAM4

Four-level pulse amplitude modulation (PAM4) encodes 2-bits into one symbol, resulting in a four-level signal and reducing the transmission bandwidth by a factor of two compared to on-off-keying. Utilizing Nyquist pulse shaping with a small roll-off factor ($\beta \approx 0.1$), the signal bandwidth can be further reduced, resulting in an electrical bandwidth of around 30 GHz for a 112 Gb/s PAM4 signal.

Fig. 6 shows the implemented offline DSP blocks for the Nyquist PAM4 system. A 4-ary deBruijn sequence of order eight ($4^8 = 65536$ symbols) is used and gray-mapped onto a PAM4 signal. Compared to DMT, PAM4 offers the possibility to easily compensate the nonlinear transfer function of the modulator by adjusting the levels towards equally spaced power levels after the modulator. Afterwards, the signal is upsampled to 3 samples/symbol, raised cosine shaped in frequency domain with $\beta = 0.1$, and downsampled by a factor of two in order to achieve an oversampling of 1.5 samples per symbols to generate a 112-Gb/s Nyquist-PAM4 signal with the 84 GS/s DAC. The subsequent digital pre-emphasis compensates the bandwidth limitations of the DAC and driver, and the signal is quantized into integer values between 0 to 255 to use the full 8-bit resolution of the DAC. Fig. 9a) illustrates the obtained eye diagrams after the driver amplifier as well as after the EML, exhibiting the typical over-and undershoots of a Nyquist PAM4 signal. Furthermore, the power spectrum density (PSD) of the transmit-signal before the DAC demonstrates the effect of pre-emphasis (gray area = uncompensated signal).

TABLE I
DMT SYSTEM PARAMETERS

| | |
|---|---|
| Modulation formats | BPSK to 64-QAM |
| Frame length (data symbols) | 124 |
| Training symbols (TS) | 4 |
| FFT length | 512 |
| Usable carriers | 255 (max. used 242) |
| Cyclic prefix (CP) | 1/64 |
| Clipping ratio | to be optimized |
| Equalizer | decision directed, 1-tap |

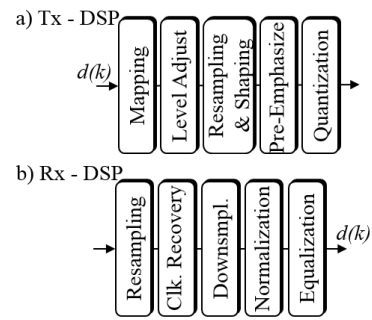

Fig. 6. DSP-flow of the implemented Nyquist PAM4 system.

At the receiver, the signal is resampled to 2 samples/symbol and for the sake of convenience an ideal clock-recovery by means of a Gardner phase detector is assumed [21]. The detector output over the phase error (S-curve) is calculated and averaged over the whole block of received data to determine the optimum phase shift. Subsequently, an adaptive symbol-spaced equalizer is applied to recover the PAM4 signal. Two different equalizer structures are evaluated: a simple adaptive feedforward equalizer (FFE) based on the least-mean square (LMS) algorithm and a combination of FFE and maximum likelihood sequence estimation (MLSE). For the latter, the MLSE basically replaces the hard decision after the FFE.

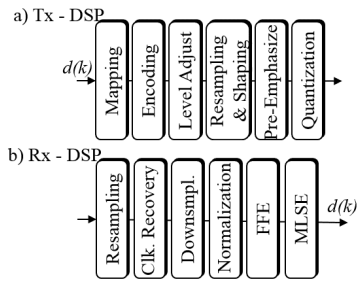

Fig. 7. DSP blocks of the implemented PR PAM4 system.

### C. Partial-Response PAM4

This paper performs partial-response encoding in the digital domain at the transmitter-side by passing the mapped PAM4 symbols $s(k)$ through a simple delay-and-add filter

$$s(k) = s(k) + s(k-1). \qquad (1)$$

This operation correlates adjacent symbols and redistributes the PSD such, that it is more concentrated at lower frequencies. Furthermore, an input PAM4 signal results in a seven-level signal at the output, which is also referred to duobinary PAM4 [11], [22]. Usually, the correlation requires pre-coding prior to the delay-and-add operation in order to avoid error-propagation at the receiver side. However, applying an MLSE at the receiver side takes the correlation into account for a decision, avoiding the need for pre-coding.

Fig. 7 shows the DSP blocks of the implemented PR PAM4 system, which resembles with some exceptions the PAM4 system. The described PR encoding is done after the PAM4 mapping and the level adjustment is now based on the obtained seven levels. Subsequently, the signal is resampled to 1.5 samples/symbols and also shaped with a Raised-Cosine filter of $\beta = 0.1$. Note, also PR PAM4 requires frequency domain pulse-shaping in order to avoid Aliasing. Again, pre-emphasis is applied and the signal is quantized before loading it into the DAC-memory. Fig. 9b) depicts the obtained PR-PAM4 eye diagrams after the driver and the EML as well as the PSD with and without pre-emphasis. The eye diagrams show clearly the seven levels due to partial response encoding.

At the receiver, the signal is first resampled to 2 samples/symbol, clock-recovery is applied and a combination of FFE and MLSE is employed to detect the symbols, both operating at one sample per symbol. Again, the FFE-coefficient calculation is based on the LMS algorithm, but with a seven-level hard decision for error estimation to recover the PR PAM4 signal.

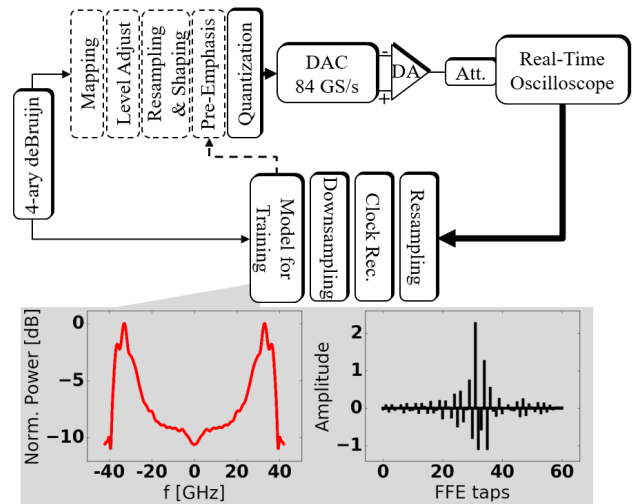

Fig. 8. Experimental setup and DSP-flow of the pre-emphasize algorithm. In the inset the obtained time domain filter coefficients are shown.

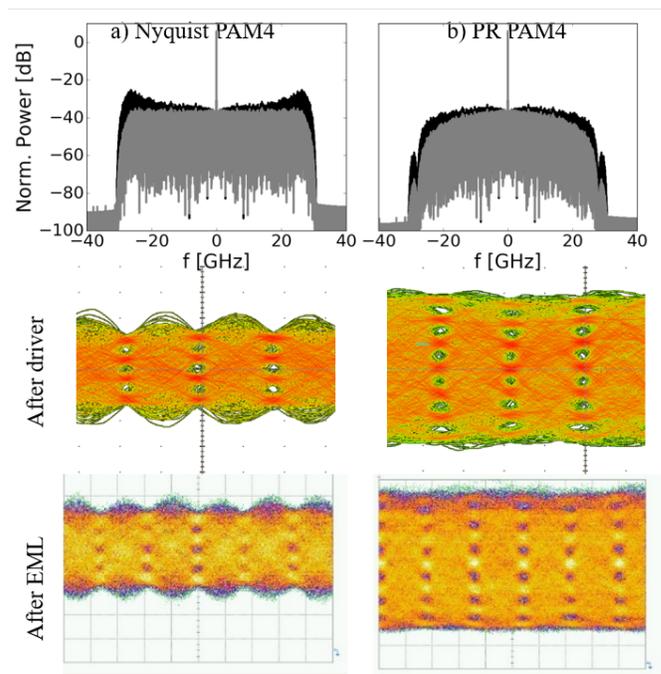

Fig. 9. Nyquist PAM4 and PR-PAM4 eye diagrams obtained after the driver amplifier and after the EML at 112 Gb/s employing digital pre-equalization. The PSDs of the transmit-signal show the effect of pre-emphasis.

## IV. DIGITAL PRE-EMPHASIS FOR SINGLE CARRIER MODULATION

To operate at 100 Gb/s and beyond in combination with higher order modulation formats, it becomes inevitable to consider the limitations of the transmitter components such as DAC, driver, electrical cables and modulator. While DMT offers the ability to do BL and PL to efficiently compensate channel distortions, single carrier modulation formats require the use of efficient pre- and post-equalization to mitigate the performance limitations at the transmitter and receiver side.

This work uses the indirect learning architecture to calculate the inverse transfer function of the transmitter components consisting of DAC, driver amplifier and the electrical cables [23], [24]. Fig. 8 depicts the experimental setup. For the estimation, a 4-ary deBruijn sequence of order eight is directly loaded onto the 84 GS/s DAC and digitized after transmission with a real-time scope of 70 GHz bandwidth and a sampling rate of 200 GS/s. The use of such high bandwidth oscilloscope ensures that only the transmitter component bandwidth distortions are taken into account. In digital domain, after resampling and clock-recovery, the pre-equalizer coefficients are calculated by comparing the received sequence with the transmitted one using the LMS algorithm. The inset of Fig. 8 displays the obtained equalizer coefficients (in this case 61 coefficients) as well as the corresponding frequency domain behaviour. Comparing the pre-emphasis for both formats, PR PAM4 requires a weaker pre-emphasis as it concentrates more energy at lower frequencies. The stronger pre-emphasis for Nyquist PAM4 with up to 10 dB amplification of the higher frequencies comes at the cost of an increased peak-to-average-power ratio (PAPR). This decreases the dynamic range of the DAC, resulting in a smaller amplitude after the DAC and driver and thus, a smaller optical modulation amplitude (OMA) and extinction ratio as demonstrated by the eye diagrams of Fig. 9 and by the optical spectrum of Fig. 17. For instance, we measured an extinction ratio (highest to lowest level) of around 4 dB for Nyquist PAM4 and an extinction ratio of around 7 dB for PR PAM4.

## V. EXPERIMENTAL RESULTS AND DISCUSSION

### A. DMT

Each format needs to be optimized individually in order to ensure a fair comparison between the different modulation formats. For DMT, several parameters such as clipping ratio, the FFT length, number of training symbols, etc. have to be optimized at first. To not go beyond the scope of this paper, some parameters with minor influence on the performance such as number of training symbols or the cyclic prefix are set to fixed values based on previous experiences/publications [25]. Fig. 10a) demonstrates the influence of different clipping ratios using an FFT length of 512, revealing an optimum of around 15 dB for this particular transmission scenario. The influence of the FFT length is shown in Fig. 10b) using a fixed clipping ratio of 15 dB and a fixed CP of 1/64 of the symbol duration. A longer FFT increases the granularity and enables a more precise adaption of the signal to the channel characteristics, resulting in an improved performance. However, the performance gain by a longer FFT decreases as only little improvement is achieved with an FFT length of 2048 compared to an FFT length of 512, whereas the implementation complexity of DMT raises proportionally to $N_{FFT}\log_2(N_{FFT})$. Hence, the use of an FFT length of 512 is a good trade-off between implementation complexity and BER-performance [25].

Fig. 11 shows the achieved BER vs. received optical power (ROP) into the PIN/TIA for different data rates as well as for different transmission distances. Since DMT allows to easily switch between different data rates by loading a different number of bits onto each subcarrier, the performance of 112 Gb/s, 89.4 Gb/s, 74.7 Gb/s and 56 Gb/s is investigated. Note that 89.4 Gb/s or 74.7 Gb/s are no Ethernet data rates and a rate of 56 Gb/s would require two wavelengths to achieve 100G transmission. Nevertheless these rates are illustrative in showing the increasing performance penalties introduced by the limited bandwidth of the system. Two different FEC thresholds are added as a solid and a dashed line, representing the standardized KP4-FEC (RS(544,514,10)) with a BER-limit of 2E-4 and the continuously-interleaved BCH FEC (CI-BCH(1020,988)) with a BER-limit of 4.4E-3 [26], [27]. Transmitting at a data rate of 112 Gb/s, BERs below the CI-BCH FEC are achieved only for the optical back-to-back (b2b) case, while BERs around the FEC threshold are achieved in case of 10 km transmission distance. At a bias of 1.25 V, the output power of the deployed EML is around 1 dBm, which results in a maximum achievable input power of $-5.5$ dBm after 20 km transmission at this wavelength ($0.32$ dB/km $* 20$ km). This input power is not sufficient to achieve BERs below the FEC-limits in case of 112 Gb/s. Indeed, a very similar performance is demonstrated for the different transmission distances up to the achievable input powers. The performance improves with decreasing bitrate.

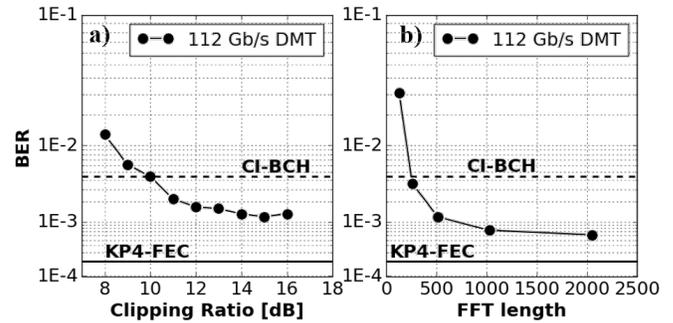

Fig. 10. Optimization of DMT parameters: a) influence of clipping ratio at an FFT-length of 512 and b) influence of FFT length at a fixed clipping ratio of 14 dB.

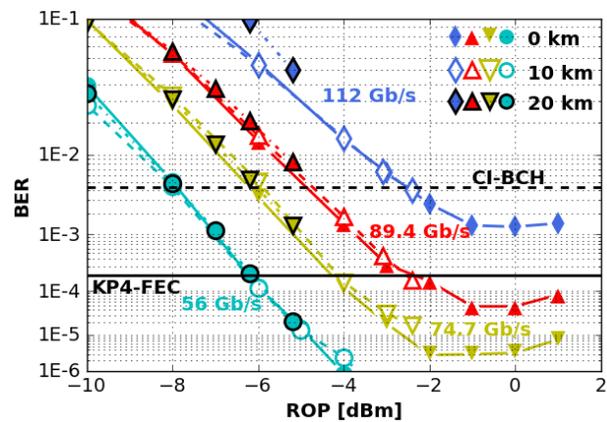

Fig. 11. Transmission results of DMT at different data rates and for different transmission distances.

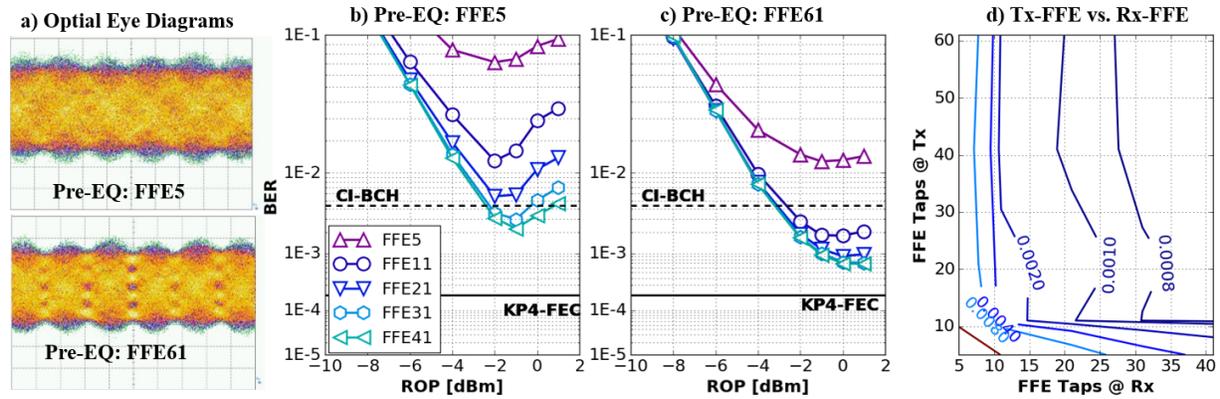

Fig. 12. Optical back-to-back transmission results of 112 Gb/s Nyquist PAM4 employing different numbers of pre- and post-FFE coefficients: a) shows the optical eye diagrams obtained directly after the EML using a pre-equalizer of 5 coefficients and 61 coefficients, b) and c) illustrate the BER vs. ROP results using different numbers of post-FFE coefficients and d) depicts the BER performance for different Tx-FFE/Rx-FFE combinations at ROP=0 dBm.

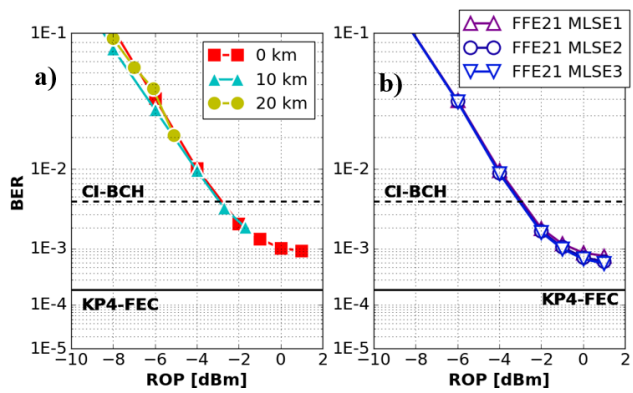

Fig. 13. Transmission results of 112 Gb/s Nyquist PAM4: a) over different transmission distances using only 11 Tx-FFE and 41 Rx-FFE coefficients and b) optical back-to-back transmission applying different MLSE memory length after the FFE.

### B. Nyquist PAM4

For Nyquist PAM4, the performance of the pre-equalizer (Tx-FFE) in combination with the applied FFE at the receiver (Rx-FFE) is evaluated in terms of BER-performance in a first step. In principle, we would like to answer the question, how many Tx-FFE and Rx-FFE coefficients are necessary for such a transmission scenario to achieve BERs below the desired FEC threshold. Fig. 12b) and c) depict the BER vs. ROP results for optical b2b, using 5 and 61 Tx-FFE coefficients, respectively, in combination with a different number of applied Rx-FFE coefficients. The number after the term "FFE" notates the number of used coefficients. Again, the previously discussed FEC thresholds are shown as black lines. The interaction between the amount of Tx-FFE and Rx-FFE coefficients is further illustrated as a contour plot in Fig. 12d), demonstrating the achieved BERs for different Tx-FFE/Rx-FFE combinations at a fixed ROP=0 dBm. Basically, no significant BER improvement is seen with more than 11 Tx-FFE coefficients, while at the receiver at least 21 Rx-FFE coefficients are required. To achieve BERs below 1E-3 however, more than 40 coefficients for both Tx-FFE and Rx-FFE are necessary.

Based on the results of Fig. 12, 11 Tx-FFE and 41 Rx-FFE coefficients offer a good tradeoff between performance and complexity and are used for further evaluation. With these settings, the performance for optical b2b, 10 km and 20 km is compared in Fig. 13a). For optical b2b and for 10 km SSMF, the results stay well below the CI-BCH FEC threshold, however, the KP4-FEC threshold is not reached. In addition, the limited output power of the EML prevents a possible transmission over 20 km. Up to the achievable input power a similar performance of the different transmission distances is shown.

Fig. 13b) illustrates the performance of different FFE-MLSE combinations for optical b2b. Again, 11 Tx-FFE coefficients are used for pre-emphasis. The number after "MLSE" indicates the used memory, e.g. an MLSE1 indicates a memory $m$ of one, which corresponds to $4^m = 4$ states for the Viterbi algorithm. Although applying a memory length of up to three, which corresponds to 64 states, only little performance improvement is achieved at higher input powers.

### C. Partial-Response PAM4

Fig. 14 shows the transmission results of PR PAM4 at 112 Gb/s. Again, the BER performance in dependence of the number of Tx-FFE and Rx-FFE coefficients is evaluated in a first step. Note, that an MLSE of memory one is employed after the Rx-FFE, in order to detect the PR PAM4 signal (see section III-C). A significant performance improvement is demonstrated also for PR PAM4 when sufficient pre-emphasis is applied. BERs below the CI-BCH FEC-limit are not achievable without any pre-emphasis. Based on the contour plot of Fig. 14c), around 11 Tx-FFE and 21 Rx-FFE coefficients are sufficient, as no further significant performance improvement is achieved. With these settings, BERs below 4E-4 can be achieved at an optimum input power of -1 dBm. At higher input powers the PIN/TIA starts to saturate, compressing the outer levels of the signal and resulting in a degradation of performance. The obtained histograms after the FFE in the inset of Fig. 14b) demonstrate this effect, as the outer levels have a smaller distance at higher ROPs.

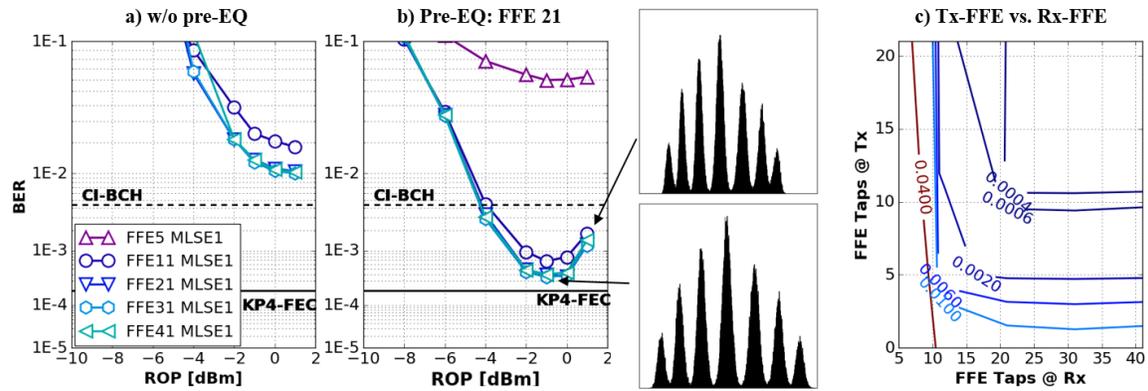

Fig. 14. Optical back-to-back transmission results of 112 Gb/s PR PAM4 employing different numbers of pre- and post-FFE coefficients: a) and b) illustrate the BER vs. ROP results using different numbers of post-FFE coefficients and d) depicts the BER performance for different Tx-FFE/Rx-FFE combinations at ROP=$-$ 1 dBm.

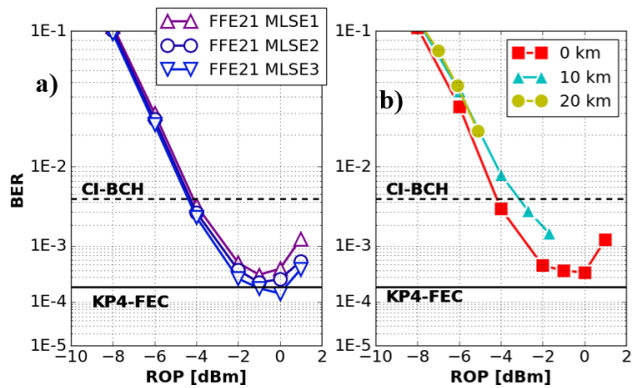

Fig. 15. Transmission results of 112 Gb/s PR PAM4: a) using different MLSE memory length after the FFE in case of optical back-to-back transmission, b) over different transmission distances using 11 Tx-FFE and 21 Rx-FFE coefficients.

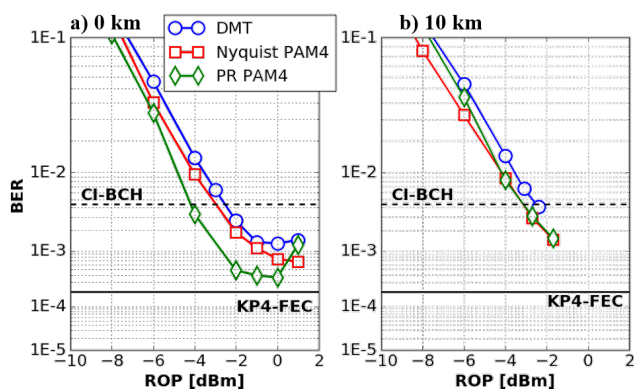

Fig. 16. Transmission results of 112 Gb/s DMT, Nyquist PAM4 and PR PAM4 for a) optical b2b and b) 10 km.

Fig. 15a) evaluates the performance using different memory length for the MLSE. Similar to Nyquist PAM4, a significant performance improvement is not obtained with a longer memory length. BERs around the KP4-FEC threshold are achieved with an MLSE3, which improves the performance from 4E-4 down to 2E-4. This "small" improvement does not justify the use of such complex equalization and thus, an MLSE1 is used to evaluate the performance up to a transmission distance of 20 km. Finally, Fig. 15b) shows the performance for b2b, 10 km and 20 km. A transmission distance of up to 10 km is possible assuming the CI-BCH FEC threshold.

*D. A comparison of DMT, Nyquist PAM4 and PR PAM4*

Finally, Fig. 16 compares the performance of 112 Gb/s DMT, Nyquist PAM4 and PR-PAM4 for optical b2b and for a transmission distance of 10 km. The optimum parameters found in the previous sections are applied for each modulation format: DMT employs an FFT length of 512, Nyquist PAM4 utilizes 11 Tx-FFE and 41 Rx-FFE coefficients and PR PAM4 uses 11 Tx-FFE and 21 Rx-FFE coefficients together with an MLSE1. For the optical b2b case, PR PAM4 shows the best transmission performance as BERs below 4E-4 are achievable. However, one has to keep in mind, that PR PAM4 uses an additional delay-and-add filter at the transmitter side and an MLSE at the receiver side, to generate and detect the partial-response signal. This is not necessary for Nyquist PAM4. DMT and Nyquist PAM4 show very similar results, only at ROPs of more than 0 dBm Nyquist PAM4 achieves lower BER values. The results reveal further, a higher sensitivity of PR PAM4 towards the nonlinear behaviour of the PIN/TIA, resulting in a significant performance degradation at higher input power values. This is also due to the fact, that the used PIN/TIA saturates faster for a signal exhibiting a higher OMA and extinction ratio. Fig. 17 shows the optical power spectrum obtained after the EML for the different modulation schemes, which on the hand reveals a significant higher extinction for PR PAM4 and on the other hand very similar spectral shapes for Nyquist PAM4 and DMT. For the transmission distance of 10 km nearly no performance differences are seen any more between the different modulation formats.

For all three modulation formats, the CI-BCH FEC is required to transmit over 10 km SSMF. The latency of this FEC is 1 Mbits, which leads to a processing delay of around 10 $\mu$s at 100 Gb/s [27]. The latency of the DSP is mainly determined by the number of equalizer taps for PAM4 or the length of the FFT for DMT. Supposing an ASIC clock of

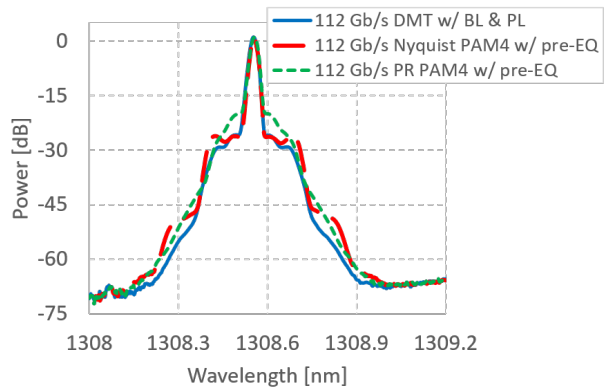

Fig. 17. Optical spectra of the different modulation formats.

1 GHz and that 56 symbols are available per ASIC clock (56 GHz/1 GHz), an FFE could process 56 symbols at the same time. The latency calculation for an $N$-tap FFE includes time for $N$ multiplications and $K=\text{ceil}(\log 2(N-1))$ additions, which would lead to a minimum latency of around 2 ns for Nyquist PAM4 with 11 Tx-FFE taps and 41 Rx-FFE taps (1 ns for Tx-FFE and 1 ns for Rx-FFE). If one multiplication and six additions cannot be performed in one clock period, e.g. when only a single operation is done in 1 ns, the worst case latency of the FFE would be 12 ns. For PR PAM4 an additional MLSE1 is required, which is here assumed to process 56 decoded symbols per single block in parallel plus 10 overhead symbols (the overhead of the MLSE is usually 5xMLSE-memory, which requires in this case 5 symbols before and after the main symbol for PR PAM4). The MLSE processing time includes one multiplication to calculate the branch metric, one addition to sum the branches and state metrics, two add-compare-select operations to select one candidate per state and one decoding update. Supposing further that a 1 GHz ASIC can process 4 symbols per clock, the MLSE1 would add around 16 ns of latency. Again, for worst case, when one clock cycle per symbol is required the MLSE1 latency is 66 ns. The overall processing delay of the DSP even in worst case is therefore negligible compared to the latency of the FEC. Estimating in addition the propagation delay with 5 $\mu$s/km, the transmission distance of 10 km leads to an overall latency of around 60 $\mu$s, which is well below the allowed one-way latency of 75 $\mu$s mentioned at the beginning.

The achieved results demonstrate at best a margin of around 2 dB with this FEC. Part of the performance penalties are due to the limited bandwidth of the electrical components (DAC, ADC and driver), which we expect to be increased in future. The EML offered sufficient bandwidth, however it exhibited a frequency dip at around 7 GHz, which unfortunately degraded the transmission performance.

## VI. CONCLUSION

We have experimentally investigated DMT, Nyquist PAM4 and partial-response PAM4, all at a data rate of 112 Gb/s as candidate modulation formats for an Ethernet-based fronthaul architecture for future 5G with 100G trunk lines between different cell sites. Employing current state-of-the-art 28 nm CMOS DAC and ADC in combination with a low-cost EML and a 25G driver, a transmission distance of 10 km SSMF is successfully bridged with all modulation formats if an HD-FEC of 4.4E-3 is assumed, allowing error free transmission. Each modulation format requires careful optimization, resulting in a very similar performance. For Nyquist PAM4 and PR PAM4 the performance depends very much on the applied pre- and post-equalization, revealing the necessity of at least 11 pre-equalizer and 21 post-equalizer FFE coefficients. On the other hand, BL and PL is essential for DMT and an FFT length of 512 is required.

## REFERENCES

[1] J. Zou, C. Wagner, and M. Eiselt, "Optical Fronthauling for 5G Mobile: A Perspective of Passive Metro WDM Technology," in *Optical Fiber Communication Conference, OFC*, Los Angeles, CA, USA, 2017, paper W4C.2.
[2] "IEEE 1914.1 Packet-based Fronthaul Transport Network," 2016. [Online]. Available: http://sites.ieee.org/sagroups-1914/p1914-1/ieee-p1914-1-draft-specifications/
[3] "IEEE 1914.3 Draft 1.2 Radio over Ethernet Encapsulations and Mappings," 2016. [Online]. Available: http://sites.ieee.org/sagroups-1914/p1914-3/ieee-p1914-3-draft-specifications/
[4] H. Jinri and Y. Yannan, "White Paper of Next Generation Fronthaul Interface," *White Paper*, 2015.
[5] Nokia, "Evolution to centralized RAN with mobile fronthaul," *Technical White Paper*, 2016.
[6] "iCirrus D3.2 Preliminary Fronthaul Architecture Proposal," 2016. [Online]. Available: http://www.icirrus-5gnet.eu/category/deliverables/
[7] M. Chagnon, M. Osman, M. Poulin, C. Latrasse, J.-F. Gagné, Y. Painchaud, C. Paquet, S. Lessard, and D. Plant, "Experimental study of 112 Gb/s short reach transmission employing PAM formats and SiP intensity modulator at 1.3 $\mu$m." *Optics Express*, vol. 22, no. 17, pp. 21 018–36, 2014.
[8] N. Kikuchi and R. Hirai, "Intensity-modulated/direct-detection (IM/DD) Nyquist pulse-amplitude modulation (PAM) signaling for 100-Gbit/s/$\lambda$ optical short-reach transmission," in *European Conference on Optical Communication, ECOC*, Cannes, France, 2014, paper P.4.12.
[9] Y. Gao, J. C. Cartledge, S. S. Yam, A. Rezania, and Y. Matsui, "112 Gb/s PAM-4 Using a Directly Modulated Laser with Linear Pre-Compensation and Nonlinear Post-Compensation," in *European Conference on Optical Communication, ECOC*, Duesseldorf, Germany, 2016, paper M2.C2.
[10] N. Stojanovic, Z. Qiang, C. Prodaniuc, and F. Karinou, "Performance and DSP complexity evaluation of a 112-Gbit/s PAM-4 transceiver employing a 25-GHz TOSA and ROSA," in *European Conference on Optical Communication (ECOC)*, Valencia, Spain, 2015, paper Tu.3.4.5.
[11] L. F. Suhr, J. J. V. Olmos, B. Mao, X. Xu, G. N. Liu, and I. T. Monroy, "112-Gbit/s x 4-Lane Duobinary-4-PAM for 400GBase," in *European Conference on Optical Communication, ECOC*, Cannes, 2014, p. Tu.4.3.2.
[12] X. Xu, E. Zhou, G. N. Liu, T. Zuo, Q. Zhong, L. Zhang, Y. Bao, X. Zhang, J. Li, and Z. Li, "Advanced modulation formats for 400-Gbps short-reach optical inter-connection," *Optics Express*, vol. 23, no. 1, p. 492, 2015.
[13] C. Xie, P. Dong, S. Randel, D. Pilori, P. Winzer, S. Spiga, B. Kögel, C. Neumeyr, and M.-C. Amann, "Single-VCSEL 100-Gb/s Short-Reach System Using Discrete Multi-Tone Modulation and Direct Detection," in *Optical Fiber Communication Conference*, Los Angeles, CA, USA, 2015, paper Tu2H.2.
[14] Y. Kai, M. Nishihara, T. Tanaka, T. Takahara, Lei Li, Zhenning Tao, Bo Liu, J. Rasmussen, and T. Drenski, "Experimental comparison of pulse amplitude modulation (PAM) and discrete multi-tone (DMT) for short-reach 400-Gbps data communication," in *European Conference on Optical Communication (ECOC)*, 2013, paper Th.1F.3.
[15] M. I. Olmedo, T. Zuo, J. B. Jensen, Q. Zhong, X. Xu, S. Popov, and I. T. Monroy, "Multiband carrierless amplitude phase modulation for high capacity optical data links," *Journal of Lightwave Technology*, vol. 32, no. 4, pp. 798–804, 2014.
[16] [Online]. Available: http://www.ieee802.org/3/bs/

[17] D. Lewis, S. Corbeil, and B. Mason, "Practical Demonstration of live-traffic Optical DMT Link using DMT Test Chip." [Online]. Available: http://www.ieee802.org/3/bs/public/14_09/lewis_3bs_01a_0914.pdf

[18] F. Caggioni, "100G single Lambda Optical link, experimental data," in *IEEE Interim Meeting - Dallas*, Sep. 2016. [Online]. Available: http://www.ieee802.org/3/cd/public/Sept16/caggioni_3cd_01_0916.pdf

[19] U. Troppenz, M. Narodovitch, C. Kottke, G. Przyrembel, W. D. Molzow, A. Sigmund, H. G. Bach, and M. Moehrle, "1.3 $\mu$m electroabsorption modulated lasers for PAM4/PAM8 single channel 100 Gb/s," in *International Conf. on Indium Phosphide and Related Materials*, Montpellier, France, 2014, paper Th.B2.5.

[20] J. M. Cioffi, "Data Transmisssion Theory: Course Text for EE379A-B and EE479," in *Multi-channel modulation*, Stanford University, Stanfort, CA, USA, 2015. [Online]. Available: http://www.stanford.edu/group/cioffi/book

[21] F. M.Gardner, *Phaselock Techniques*. Wiley-Interscience, 1980.

[22] S. Walklin and J. Conradi, "Multilevel signaling for increasing the reach of 10 Gb/s lightwave systems," *Journal of Lightwave Technology*, vol. 17, no. 11, pp. 2235–2248, 1999.

[23] G. Khanna, B. Spinnler, S. Calabrò, E. De Man, and N. Hanik, "A Robust Adaptive Pre-Distortion Method for Optical Communication Transmitters," *IEEE Photonics Technology Letters*, vol. 28, no. 7, pp. 752–755, 2016.

[24] C. Eun and E. J. Powers, "A new Volterra predistorter based on the indirect learning architecture," *IEEE Transactions on Signal Processing*, vol. 45, no. 1, pp. 223–227, 1997.

[25] A. Dochhan, H. Griesser, N. Eiselt, M. Eiselt, and J.-P. Elbers, "Optimizing Discrete Multi-tone Transmission for 400G Data Center Interconnects Results an Discussion Launch power," *ITG Symp. Photonic Networks*, 2016.

[26] "IEEE P802.3bs/D1.2 Draft Standard for Ethernet Amendment: Media Access Control Parameters, Physical Layers and Management Parameters for 400 Gb/s Operation," pp. 1–269, 2016. [Online]. Available: http://www.ieee802.org/3/bs/

[27] M. Scholten, T. Coe, and J. Dillard, "Continuously-interleaved BCH (CI-BCH) FEC delivers best in class NECG for 40G and 100G metro applications," in *2010 Conference on Optical Fiber Communication (OFC/NFOEC), collocated National Fiber Optic Engineers Conference*, 2010, paper NTuB3.